\documentclass[10pt]{article}

\usepackage{amsmath, hyperref}
\usepackage{amssymb}
\usepackage{mathrsfs}
\usepackage{lmodern}
\usepackage{graphicx}
\usepackage{subcaption}
\usepackage{braket}
\usepackage{lipsum}
\usepackage{float}
\usepackage{cite}
\usepackage[left=2.5cm,right=2.5cm,top=1.2cm,bottom=1.3cm,includeheadfoot]{geometry}
\usepackage{indentfirst}
\usepackage{color}

\newcommand{\vect}[1]{\textbf{#1}}
\newcommand{\mtica}[1]{\textbf{\texttt{#1}}}

\begin{document}

\begin{center}
{\Large\bf Relativistic effects on the Schr\"odinger-Newton equation}
\vskip 4mm

David Brizuela
\footnote{
E-mail address: david.brizuela@ehu.eus}
and
Albert Duran-Cabacés
\footnote{ E-mail address: albertdurancabaces@gmail.com}

\vskip 2mm
{\sl Department of Physics and EHU Quantum Center, University of the Basque Country UPV/EHU,\\
	Barrio Sarriena s/n, 48940 Leioa, Spain}

\end{center}
\begin{quotation}
\centerline{\textbf{Abstract}}

\noindent

The Schr\"odinger-Newton model describes self-gravitating quantum particles, and it is often
cited to explain the gravitational collapse of the wave function and the localization of macroscopic objects.
However, this model is completely nonrelativistic. Thus, in order to study whether the relativistic effects may
spoil the properties of this system,
we derive a modification of the Schr\"odinger-Newton equation
by considering certain relativistic corrections
up to the first post-Newtonian order. The construction of the model begins by considering the Hamiltonian of a relativistic particle propagating
on a curved background. For simplicity, the background metric is assumed to be spherically symmetric
and it is then expanded up to the first post-Newtonian order. After performing the canonical quantization of the system,
and following the usual interpretation, the square of the module of the wave function defines a mass distribution,
which in turn is the source of the Poisson equation for the gravitational potential.
As in the nonrelativistic case,
this construction couples the Poisson and the Schr\"odinger equations and leads to a complicated nonlinear system.
Hence, the dynamics of an initial Gaussian wave packet is then numerically analyzed.
We observe that the natural dispersion of the wave function is slower
than in the nonrelativistic case. Furthermore, for those cases that reach a final localized stationary state,
the peak of the wave function happens to be located at a smaller radius.
Therefore, the relativistic corrections effectively contribute to increase the self-gravitation of the particle
and strengthen the validity of this model as an explanation for the gravitational localization of the wave function.

\end{quotation}

\vskip 2mm

\noindent

\section{Introduction}

The Schr\"odinger-Newton (SN) equation \cite{Bahrami2014} describes nonrelativistic quantum objects under self-gravitation.
In this model, the Schrödinger equation is coupled to a Newtonian gravitational potential term, which, in turn, is sourced
by the mass density given by the square of the module of the wave function. Therefore, contrary to the usual
Schr\"odinger equation, this system is nonlinear. The SN equation was first proposed to study self-gravitating bosonic
stars \cite{Ruffini1969}. However, interesting applications of this equation 
to describe the gravitational collapse of the wave function and the decoherence of macroscopic objects
have also been proposed later \cite{Diosi1984, Penrose1996}.

Despite its nonlinearity, the SN system has been shown to have an infinite family of defined stationary states with negative energies \cite{Moroz1998}. These eigenstates have been explicitly found both
numerically \cite{Bernstein1998, Moroz1998, Tod2001, Harrison2002} and with approximate analytical methods \cite{Mak2021}. The studies of the dynamical evolution of a general wave packet given by the SN equation
have been mostly based on numerical methods due to the complexity of the nonlinear system (see, e.g., Refs. \cite{VanMeter2011, Harrison2002, Guzman2004, Guzman2006, Giulini2011, Manfredi2013, Silvestrini2015}).
In general, in this dynamical scenario, two regimes can be
distinguished depending on the features of the initial state. On the one hand, there is the weak self-gravitational regime,
where the natural quantum spreading of the wave function dominates. In this regime, the wave function evolves similarly
to, but somehow slower than, a free particle. On the other hand, there is the case dominated by self-gravitation, or, as we will refer to it,
the strong self-gravitational regime.
In this latter regime, the wave function tends to decay into its corresponding ground state
through the so-called gravitational cooling \cite{Guzman2004, Guzman2006}.

Furthermore, certain modifications of the SN equation have also been studied in the literature.
For instance, a dark-energy term was coupled to this equation in Refs. \cite{Kelvin2019, Mak2021}.
The addition of dark energy changes the eigenstates of the Hamiltonian,
as well as the dynamical evolution of
the wave function, but the main features of the model are kept intact.
 
All in all, the SN equation is a promising effective model to describe macroscopic quantum objects,
which are expected to be well localized. Nonetheless, due to the Newtonian treatment of the gravitational interaction,
relativistic effects,
which could strengthen or undermine the validity of this model, are completely missing from this
picture. In fact,
the Newtonian limit of fully relativistic self-gravitating models, like the Einstein-Klein-Gordon system,
has been analyzed \cite{Guzman2004}, and certain variations of the Klein-Gordon equation with a Newtonian gravitational
potential have been proposed \cite{Mendonsa2019}.
In addition, there are studies in which relativistic and gravitational corrections from external fields are considered
in the usual Schrödinger equation \cite{Schwartz2019, Wajima1997}, though discarding the self-gravitational interaction.
In Ref. \cite{Manfredi15}, the SN equation was derived in
the context of the gravitoelectromagnetism approximation to general relativity.
While this approximation includes the relativistic corrections produced by
the gravitomagnetic vector potential up to first post-Newtonian (1PN) order (that is,
up to order $c^{-2}$, with $c$ being the speed of light), other terms of the same
order are excluded.
Therefore, the main goal of the present paper is to construct a
model that describes the SN system with the 1PN relativistic
corrections given by nonlinear terms of the Newtonian potential and kinetic energies,
which have not been previously considered,
and to determine the physical consequences of such corrections on the dynamics of the wave packets.

For such a purpose, we will first derive the Hamiltonian of a relativistic particle propagating
on a curved background. After assuming spherical symmetry, and performing a post-Newtonian
expansion of the metric in inverse powers of the speed of light up to the order $c^{-2}$, the basic
variables will be promoted to operators, and the canonical
quantization of the Hamiltonian will be performed. In this way, we will obtain the Schr\"odinger
equation with relativistic corrections, which will depend on a certain Newtonian gravitational potential.
This potential will then be assumed to obey the Poisson equation with a mass density given
by the probability distribution of the particle. This procedure will lead to a closed nonlinear
system of equations. We will then analyze the dynamics of an initial Gaussian state given by
these equations of motion, and obtain
the physical effects produced by the relativistic correction terms by comparing this evolution
with the one given by the usual nonrelativistic SN system.

The paper is organized as follows. In Sec. 2, the SN equation with
relativistic corrections up to the first post-Newtonian order is derived.
Section 3 presents the analysis of the dynamics of an initial Gaussian wave packet.
In Sec. 4 we summarize and discuss the main physical results of the model.
Finally, in Appendix A we include a description and some technical aspects
of the numerical methods used to solve the equations,
while in Appendix B we present the results found with a modified version
of the Poisson equation.

\section{Relativistic corrections on the Schr\"odinger-Newton equation}

This section is divided into three subsections. In Sec. \ref{sec_hamiltonian} the Hamiltonian
of a relativistic particle propagating on a certain curved background is obtained. In Sec. \ref{sec_PN}
a post-Newtonian expansion of the Hamiltonian is considered up to 1PN order.
Finally, in Sec. \ref{sec_quantization}, the canonical quantization of the system is performed in order to obtain
the Schr\"odinger-Newton equation with relativistic corrections.

\subsection{Hamiltonian of a relativistic particle on curved backgrounds}\label{sec_hamiltonian}

The action of a particle with mass $m$
propagating on a spacetime described by the metric tensor $g_{\mu\nu}$
is given by
\begin{equation}
 S=\int ds\, L(x^\mu,\dot{x}^\mu),
\end{equation}
with the Lagrangian

\begin{equation}
     L(x^\mu,\dot{x}^\mu)=-mc(- g_{\mu \nu} \dot{x}^{\mu}\dot{x}^{\nu})^{{1/2}},
    \label{eq_lagrangian}
\end{equation}
where $x^\mu=x^\mu(s)$ is the trajectory of the particle, and
the dot stands for a derivative with respect to the parameter $s$.
In this setup, all four coordinates $x^\mu=(x^0,x^1,x^2,x^3)$ are dynamical variables. For convenience, we will also use the notation
$x^0:=ct$ for the time coordinate.
In order to obtain the Hamiltonian, one first needs to define the conjugate momenta,
\begin{equation}
 p_\mu:=\frac{\partial L}{\partial\dot{x}^\mu}=\frac{mc}{(-g_{\alpha\beta}\dot{x}^{\alpha}\dot{x}^{\beta})^{1/2}}g_{\mu\nu}\dot{x}^\nu.
\end{equation}
It is easy to check that these momenta are not independent, since they obey
the constraint
\begin{equation}\label{constraint}
 p_\mu p^\mu+m^2c^2=0.
\end{equation}
This constraint is first class, and it is related to the reparametrization invariance
of the system, that is, the freedom to choose the parameter $s$.
By performing a Legendre transformation and following the usual Dirac procedure
for constrained systems, which states that one should add the different constraints
multiplied by certain Lagrange multiplier to the Hamiltonian,
one then finds the generalized Hamiltonian,
\begin{equation}\label{generalizedh}
 {\cal C}= \frac{\alpha}{2} ( p_\mu p^\mu+m^2c^2),
\end{equation}
with the Lagrange multiplier $\alpha$. The equations of motion can be readily obtained
by computing the Poisson brackets between the different variables and the Hamiltonian,
\begin{align}
\dot{x}^\mu&=\{x^\mu,{\cal C}\}=\alpha g^{\mu\nu} p_\nu,\\
\dot{p}_\mu&=\{p_\mu,{\cal C}\}=-\frac{\alpha}{2}p_\alpha p_\beta \frac{\partial g^{\alpha\beta}}{\partial x^{\mu}}.
\end{align}
Note that the Hamiltonian \eqref{generalizedh} is vanishing on shell and its canonical quantization
would lead to the Klein-Gordon equation. Instead, in order to construct our model,
we will fix the gauge before quantization by imposing the condition
$x^0=c s$, or equivalently $t=s$, on the time coordinate.
The conservation of this condition all along the evolution, $\dot{x}^0=c$, establishes the value $\alpha=c/(g^{0\mu}p_\mu)$
for the Lagrange multiplier.
The gauge-fixing procedure is then completed by solving the constraint \eqref{constraint}
to write the conjugate momentum $p_0$ in terms of the other variables. This leads to the Hamiltonian
\begin{equation}
    H:=-cp_0=\frac{c}{\sqrt{-g^{00}}}\left[m^2 c^2+\left(g^{ij}-\frac{1}{g^{00}}g^{0i}g^{0j}\right)p_i p_j\right]^{{1/2}}+c \frac{ g^{0i}}{g^{00}}p_i,
    \label{eq_hamiltonian}
\end{equation}
where we have chosen the positive sign in front of the square root in order
to get the correct sign in the nonrelativistic limit and latin
letters stand for spatial indices running from 1 to 3. Now, the dynamical
variables are the three spatial coordinates $x^i$, for which we will also use
the notation ${\bf x}:=(x^1,x^2,x^3)$, and the equations of motion take the form
\begin{align}
  \dot{x}^i&=c\, (g^{00}H-c g^{0k}p_k)^{-1}(g^{0i}H-c g^{ij}p_j),\\
 \dot{p}_i&=\frac{1}{2}\left(g^{00}H-c g^{0k}p_k\right)^{-1}\left(H^2\frac{\partial g^{00}}{\partial x^i}
 -2 c H p_j \frac{\partial g^{0j}}{\partial x^i}+c^2p_jp_l \frac{\partial g^{jl}}{\partial x^i} \right).
 \end{align}

\subsection{Parametrized post-Newtonian formalism}\label{sec_PN}

In the nonrelativistic limit,
every metric theory of gravity can be written in terms of small deviations from the Newtonian
gravitational equations.
The post-Newtonian framework encompasses the mathematical tools used for this purpose.
An especially useful setup is the so-called parametrized post-Newtonian formalism
\cite{Will2018}, which encodes these deviations using explicit parameters. These parameters have specific physical meaning, as they describe different properties
of the spacetime, and their numerical values are fixed by the particular
theory of gravity under consideration.

However, there is nothing like a parametrized post-Newtonian formalism for a generic spacetime. In order to construct
such a framework, one needs
to assume certain particular physical scenario with corresponding symmetries and, more importantly,
a specific matter content. The usual parametrized post-Newtonian formalism is written for a perfect fluid, which
is a suitable assumption for most applications,
and it is described in terms of ten real-valued parameters along with the same
number of metric potentials that obey Poisson-like equations.
This formalism includes all possible 1PN-order terms in the metric.
However, the resolution of such a system is very complicated due to the large
number of equations involved. In addition, since our goal is to describe
the evolution of a mass density given by the norm of the wave function,
in order to apply the complete formalism, one would need to define different
kinematic and thermodynamic quantities (like the velocity, the pressure,
the internal energy,...) in terms of the wave function,
as well as the equation of state of such a fluid, and such definitions
might not be obvious. Therefore, we leave the construction and analysis of the complete framework
for future work. Here,
in order to construct the simplest possible model that encodes
the 1PN corrections that only involve the Newtonian potential,
we will assume the same metric as considered by Eddington, Robertson, and Schiff \cite{Will2018},
which describes a spherically symmetric vacuum spacetime.
By doing so, we are neglecting all other post-Newtonian potentials.
Some of these potentials are related to the velocity of the fluid, while
others are sourced by thermodynamic magnitudes as the pressure or the internal
energy. Therefore, our formalism will be valid near the equilibrium and
as long as such termodynamic quantities do not have a large impact on the dynamics.
Finally, let us comment that, instead of using the parametrized post-Newtonian framework
to study the relativistic effects,
another alternative would be to consider the metric obtained from the expansion of the
Einstein-Klein-Gordon system presented in Ref. \cite{Giulini12}, which leads to the SN system.

In the coordinates centered at the gravitational source, the components of the
Eddington-Robertson-Schiff metric are
\begin{equation}
   g_{00}=-\left(1+2\frac{\Phi}{c^2}+2\beta\frac{\Phi^2}{c^4}\right)+\mathcal{O}(c^{-5}), \quad g_{0i}={\cal O}(c^{-4}), \quad g_{ij}=\left(1-2\gamma\frac{\Phi}{c^2}\right)q_{ij}+\mathcal{O}(c^{-3}),
    \label{eq_metricppn}
\end{equation}
where $q_{ij}$ is the three-dimensional Euclidean metric,
$\Phi$ is the Newtonian gravitational potential, and
the truncation order is chosen so that all the components of the line element $g_{\mu \nu}dx^{\mu}dx^{\nu}$ are accurate up to an order $c^{-2}$.
In particular, this differs from the gravitoelectromagnetism approximation
used in Ref. \cite{Manfredi15}, which does not include the quadratic term $\Phi^2$ in
the expansion of the $g_{00}$ component of the metric.
Note that this metric depends on the two parameters $\beta$ and $\gamma$. The parameter $\beta$
describes how nonlinear the superposition law for gravitational fields is, whereas
$\gamma$ is related to the spatial curvature. Both of them have fixed values for any given theory
of gravity and, in the particular case of general relativity, $\beta=\gamma=1$.
However, in order to track their contribution along the different equations, we will keep them explicitly
and only replace them by their value in general relativity $\beta=\gamma=1$
for the numerical implementation that will be performed in the next section.

It is now straightforward to obtain
the components of the inverse metric,
\begin{equation}
    g^{00}=-\left(1-2\frac{\Phi}{c^2}-(2\beta-4)\frac{\Phi^2}{c^4}\right)+\mathcal{O}(c^{-5}),
     \quad g^{0i}={\cal O}(c^{-4}),\quad \quad g^{ij}=\left(1+2\gamma\frac{\Phi}{c^2}\right)q^{ij}+\mathcal{O}(c^{-3}),
    \label{eq_invmetricppn}
\end{equation}
with $q^{ij}$ being the inverse Euclidean metric, that is, $q^{ij}q_{jk}=\delta^i{}_k$.
Replacing the elements \eqref{eq_invmetricppn}
into the expression
\eqref{eq_hamiltonian}, and performing an expansion in inverse powers of $c$,
the Hamiltonian up to 1PN order is found to be
\begin{equation}
    H=m c^2+\frac{\vect{p}^2}{2m}+m\Phi+\frac{m}{2c^2}(2\beta-1)\Phi^2+\frac{(2\gamma+1)}{2mc^2}\Phi\vect{p}^2-\frac{\vect{p}^4}{8m^3c^2}+\mathcal{O}(c^{-3}),
    \label{eq_HamiClass}
\end{equation}
with $\vect{p}^2:=p_ip_jq^{ij}$. The first term is the rest energy of the particle and
will be absorbed simply by redefining the origin of the energy. The second and third terms
are, respectively, the Newtonian kinetic and potential energies. The last three terms
are thus the relativistic corrections, which are essentially the three possible quadratic
combinations of the Newtonian kinetic and potential energies. That is, the term that depends
on $\beta$ is quadratic on the potential energy and, as commented above, represents how
nonlinear the superposition of potential energies is, which is exactly linear in the Newtonian regime.
The term parametrized by $\gamma$ shows a coupling between the potential and kinetic
energies, while the last term is quadratic in the kinetic energy and encodes
the usual relativistic correction to the kinetic energy.
These relativistic terms are not present in the
Hamiltonian obtained in the context of the gravitoelectromagnetism approach (see equation (30) of Ref. \cite{Manfredi15}), while in contrast, the geometromagnetic
potential vector is absent from our Hamiltonian since it corresponds
to one of the neglected post-Newtonian potentials.

\subsection{Canonical quantization}\label{sec_quantization}

In order to perform the canonical quantization of the system, the classical position $x^i$ and momentum $p_i$ variables
are promoted to quantum operators $x^i\rightarrow \hat{x}^i$ and $p_i \rightarrow \hat{p}_i$.
These operators do not commute,
\begin{equation}
    [\hat{x}^i,\hat{p}_j]=i\hbar\delta^i{}_{j},
    \label{eq_commutation}
\end{equation}
and they act on states of a given Hilbert space $\mathcal{H}$.
For any two vectors $\varphi,\phi\in {\cal H}$,
the inner product of this space will be defined as

\begin{equation}
    \braket{\varphi,\phi}=\int_{\mathbb{R}^3}  \overline{\varphi(\vect{x},t)}\phi(\vect{x},t) \sqrt{q}\,d^3x,
    \label{eq_innerprod}
\end{equation}
with the overline denoting the complex conjugate and $q$ being the determinant of the three-dimensional Euclidean metric.
Note that, in curved spacetimes, instead of the `flat' measure $\sqrt{q}\,d^3x$, one sometimes considers the
volume element $\sqrt{g}\,d^3x$, with $g$ being the determinant of the spatial metric $g_{ij}$, to define the corresponding inner product.
If the metric $g_{ij}$ is time independent, both quantization schemes will lead to unitary
equivalent theories.
Nonetheless, if the metric $g_{ij}$ is time dependent, the inner product given by $\sqrt{g}\,d^3x$
will also be time dependent and thus the evolution of the wave function as given by
the corresponding Schr\"odinger equation will not automatically define an isometry
between Hilbert spaces.
Therefore, it turns out to be more convenient to use the flat inner product as defined above.

In this representation, the position operator $\hat x^i$ will act as a multiplication operator,
that is, $\hat x^i\varphi:=x^i\varphi$, whereas the
action of the momentum
operator will be defined as $\hat p_j\varphi:=-i\hbar q^{-1/4}\partial_j (q^{1/4}\varphi)$
for any $\varphi\in{\cal H}$.
These operators obey the above commutation relations \eqref{eq_commutation} and are both
Hermitian under the inner product \eqref{eq_innerprod}.
Apart from this, in order to obtain the explicit action of the quantum Hamiltonian operator,
one needs to decide the ordering of the
basic operators in the different terms of the expression \eqref{eq_HamiClass}.
In particular, the terms $\vect{p}^2$, $\Phi \vect{p}^2$, and $\vect{p}^4$ are sensitive
to ordering ambiguities since in those terms position and momenta variables appear coupled.
The Hermiticity of the Hamiltonian requires a symmetric ordering of the basic operators.
Therefore, for our model, we will
choose, on the one hand, ${\bf p}^2$ to be written as $q^{-1/4}p_i q^{1/4} q^{ij}q^{1/4}p_i q^{-1/4}$
so that its quantization leads to the covariant expression
$\hat {\bf p}^2:=-\hbar^2\Delta=-\hbar^2 q^{ij}\nabla_i\nabla_j$, with $\Delta$
being the flat Laplacian operator and $\nabla_i$ being the covariant derivative associated with
the Euclidean metric $q_{ij}$. In addition, $\vect{p}^4$ will simply be understood as $(\vect{p}^2)^2$.
On the other hand, the term $\Phi \vect{p}^2$ will be quantized as $(\hat{\vect{p}}^2\Phi+\Phi\hat{\vect{p}}^2)/2$. This is certainly an
arbitrary choice, but it has a nice feature: as shown in Ref. \cite{Schwartz2019}, the Hamiltonian
operator obtained by the present canonical quantization is equivalent to the one found by considering
the nonrelativistic limit of the Einstein-Klein-Gordon equation. In this way, the Hamiltonian operator takes thus the explicit form
\begin{align}
    \hat{H}=-\frac{\hbar^2}{2m}\Delta+m\Phi(\vect{x})+\frac{m(2\beta-1)}{2c^2}\Phi^2(\vect{x})
    -\frac{\hbar^2(2\gamma+1)}{4mc^2}( \Phi(\vect{x})\Delta+ \Delta \Phi(\vect{x}))-\frac{\hbar^4}{8m^3c^2}\Delta \Delta,
    \label{eq_hamoprepr}
\end{align}
and the dynamics of the wave function
$\psi=\psi(t, \vect{x})$
is ruled by the Schr\"odinger equation,
\begin{align}
    i\hbar \frac{\partial\psi}{\partial t}&=\hat H\psi.
    \label{eq_snrc}
\end{align}

Now, in order to solve this equation, one needs to provide the Newtonian potential $\Phi$.
In our case, we will assume that the source of this potential is the particle itself, since,
due to its quantum character, its exact position is not well defined,
but instead, it is given in terms of the probability
distribution $|\psi(\vect{x},t)|^2$. Therefore, following this interpretation,
$m |\psi(\vect{x},t)|^2$ defines a mass density, which
in turn produces the gravitational potential $\Phi$ as given by the Poisson equation,
\begin{equation}
\Delta \Phi=4 \pi G m |\psi|^2.
\label{eq_snrcpoisson}
\end{equation}
Hence, the system of equations \eqref{eq_snrc}--\eqref{eq_snrcpoisson} is closed, and
it describes the evolution of a self-gravitating spherically symmetric quantum particle,
taking into account the relativistic effects given by the Newtonian potential up to an order $c^{-2}$.
However, following the basic principles of relativity,
one also would expect that not only the mass density, but also
the energy density of the gravitational field should source the Poisson
equation. In particular, in Refs. \cite{Franklin15, Franklin16} it was argued that a
consistent self-coupling of the energy density
leads to the equation $\Delta(\sqrt{\Phi+c^2})=\frac{2\pi G\rho}{c^2} \sqrt{\Phi+c^2}$.
Expanding this expression in inverse powers of $c$, one gets the modified
form of the Poisson equation $\Delta\Phi=4\pi G\rho(1+\frac{\Phi}{c^2})+\frac{(\nabla\Phi)^2}{2 c^2}$
at 1PN order. For the numerical resolution of the system, we have also considered
this form of the Poisson equation and find out that the qualitative results do
not differ much from the ones obtained with equation \eqref{eq_snrcpoisson}.
Therefore, from now on we will consider that the potential $\Phi$ obeys
\eqref{eq_snrcpoisson}. The subtle differences introduced by the modified
Poisson equation are commented on in Appendix B.

Due to the symmetry of the system, it is natural to choose spherical coordinates
in such a way that all the functions only depend on the radius $r$ and time $t$, i.e.,
$\psi=\psi(r,t)$
and $\Phi=\Phi(r,t)$. In these coordinates, $\sqrt{q}=r^2\sin\theta$ and the inner product \eqref{eq_innerprod}
between two vectors of the Hilbert space, $\varphi$ and $\phi$, takes the form
\begin{equation}
 \langle \varphi,\phi \rangle=4\pi\int_0^\infty \overline{\varphi(r,t)}\phi(r,t)r^2 dr.
\end{equation}
Thus, for convenience, we will define the radial wave function $\chi:=2\sqrt{\pi}\,r\psi$, with
its normalization simply given by
$\int_0^\infty|\chi|^2 dr=\langle\psi,\psi\rangle=1$
and with a clear physical interpretation: $|\chi|^2dr$ provides the probability to find
the particle between a radius $r$ and $r+dr$.
In this way, the Schr\"odinger-Newton (SN) system \eqref{eq_snrc}--\eqref{eq_snrcpoisson}
up to 1PN order takes the explicit form
\begin{align}
& i\hbar \frac{\partial\chi}{\partial t}=\!-\frac{\hbar^2}{2m}\frac{\partial^2\chi}{\partial r^2}+\! m\Phi\chi-\!\frac{\hbar^4}{8m^3c^2} \frac{\partial^4\chi}{\partial r^4}+\frac{m(2\beta-1)}{2c^2}\Phi^2\chi
-\!\frac{\hbar^2(2\gamma+1)}{4mc^2}\left( \frac{\partial^2\Phi}{\partial r^2} \chi+2 \frac{\partial \Phi}{\partial r} \frac{\partial \chi}{\partial r} +2\Phi\frac{\partial^2\chi}{\partial r^2}\right),
    \label{eq_SNRCspherical}
   \\
   &\frac{\partial}{\partial r}\left(r^2\frac{\partial\Phi}{\partial r}\right)= G m |\chi|^2.
\label{eq_PoissonSpherical}
\end{align}
It is straightforward to check that in the limit $c\rightarrow\infty$ all relativistic correction
terms vanish and one recovers the usual
SN system in spherical symmetry:
\begin{align}
& i\hbar \frac{\partial\chi}{\partial t}=\!-\frac{\hbar^2}{2m}\frac{\partial^2\chi}{\partial r^2}+\! m\Phi\chi,
    \label{eq_SNRCspherical0}
   \\
   &\frac{\partial}{\partial r}\left(r^2\frac{\partial\Phi}{\partial r}\right)= G m |\chi|^2.
\label{eq_PoissonSpherical0}
\end{align}

The main goal of the following section is to check
the physical effects of the relativistic correction terms. However,
the above equations are highly coupled, and it is very difficult to obtain any analytical
information from them. Therefore, for such a purpose, we will consider certain initial
wave packet and numerically check the differences between the evolution
provided by the 1PN-relativistic \eqref{eq_SNRCspherical}--\eqref{eq_PoissonSpherical} and the
nonrelativistic \eqref{eq_SNRCspherical0}--\eqref{eq_PoissonSpherical0} SN systems.
Since we are assuming that the relativistic corrections are small, they will be interpreted as
perturbations. Therefore, we will also refer to the equations \eqref{eq_SNRCspherical}--\eqref{eq_PoissonSpherical} as the perturbed system and to the equations 
\eqref{eq_SNRCspherical0}--\eqref{eq_PoissonSpherical0} as the background system.
In this respect, it is important to note that the perturbed system is of higher order
in radial derivatives than the background one. Therefore, this is a singular perturbation problem,
and there might be solutions to the equations \eqref{eq_SNRCspherical}--\eqref{eq_PoissonSpherical}
that do not tend to a solution of the background system in the limit $c\rightarrow\infty$.
As will be discussed below, this kind of solution will be avoided by imposing appropriate
boundary conditions to the perturbed system that are obeyed in a natural way by all the solutions of the background equations.

\section{Evolution of a Gaussian wave packet}

This section is divided into four subsections. In Sec. \ref{sec_initialboundary} we present
the initial Gaussian state and the boundary conditions to be considered for the numerical integration.
In Sec. \ref{sec_dimensionless} a dimensionless version of the system is constructed, so that
the number of parameters is reduced to the minimum required. Sec. \ref{sec_validity}
then discusses the possible ranges that can be chosen for those parameters
and defines the set of cases that will be studied in order to analyze in an efficient way a large section of the parameter space.
Finally, in Sec. \ref{sec_results} the obtained physical results are discussed.

\subsection{Initial and boundary conditions}\label{sec_initialboundary}

For the numerical study, we will choose a
spherical Gaussian wave packet centered at the origin
as the initial state for the wave function $\psi$. This choice
is taken due to its spherical symmetry and its usual interpretation in
quantum mechanics as the minimal uncertainty state, in the sense that
it saturates the Heisenberg relation with the same uncertainty
for both position and momentum variables. Furthermore,
several studies in the literature (see, e.g., Refs. \cite{Guzman2004, Harrison2002, VanMeter2011, Giulini2011, Guzman2006, Silvestrini2015, Manfredi2013})
have also used an initial spherical Gaussian wave packet
to study the SN equation numerically,
which will allow us to perform a direct comparison with the results obtained in the present work.
Therefore, the initial profile of the radial wave function reads,
\begin{equation}
    \chi(r,0)=2\sqrt{\pi}r\psi(r,0)=\frac{2\,r}{(\pi \sigma^{6})^{1/4}}\,e^{- \frac{r^2}{2\sigma^2}},
    \label{eq_initialstate}
\end{equation}
with $\sigma$ being the standard deviation of the initial state, which will also be referred to as the Gaussian width. 
Note that the peak (the maximum) of this function is located at $r=\sigma$,
which defines the mode of the initial probability distribution $|\chi|^2 dr$, and thus
the most probable position of the particle. Once this initial state is chosen,
the initial form of the gravitational potential $\Phi(r,0)$ is not free, and one
can find it by solving
the elliptic equation \eqref{eq_PoissonSpherical},
\begin{equation}
 \Phi(r,0)=c_1+\frac{c_2}{r}+\frac{G m}{r}{\rm erf}(r/\sigma),
\end{equation}
where $c_1$ and $c_2$ are integration constants, and ${\rm erf}$ is the error function.

Concerning the boundary conditions, first of all,
to avoid irregularities at the origin and to ensure a finite norm of the wave function, the radial wave function $\chi$
has to vanish both at the origin and at infinity, i.e., $\chi(r=0,t)=0=\chi(r=\infty,t)$.
In a similar way, the gravitational potential needs to be regular at the origin, which is translated to the condition $\partial_r\Phi(r,t)|_{r=0}=0$ and implies $c_2=0$. In addition, in order to obtain a Newtonian behavior at long distances, the Robin boundary condition $\partial_r \Phi(r,t)=-\Phi(r,t)/r$ will be
imposed at $r\rightarrow \infty$, and thus $c_1=0$.
These four boundary conditions are enough to solve the background system \eqref{eq_SNRCspherical0}--\eqref{eq_PoissonSpherical0}, but
since the perturbed SN equation \eqref{eq_SNRCspherical} contains up to fourth-order radial derivatives
of $\chi$, to solve the perturbed system \eqref{eq_SNRCspherical}--\eqref{eq_PoissonSpherical}, one needs to impose two further boundary
conditions on $\chi$.
In particular, as already commented above,
we will choose these two new boundary conditions
for the perturbed problem as relations that are automatically satisfied by the background system.
In this way, we will obtain a solution that can be considered a perturbation
in the sense that it will tend to a background solution as $c\rightarrow\infty$.
More precisely, taking the limit $r\rightarrow 0$ on the background SN equation
\eqref{eq_SNRCspherical0}, along
with the condition $\chi(r=0,t)=0$, leads to the identity $\partial^2_r\chi(r,t)|_{r=0}=0$,
which is obeyed by all background solutions. 
Furthermore,
applying the second derivative with respect to $r$ to both sides of the background SN equation
\eqref{eq_SNRCspherical0}, and considering the above conditions for the radial
wave function and gravitational potential at the origin, the relation $\partial^4_r\chi(r,t)|_{r=0}=0$ is obtained. Therefore, the two further boundary conditions to implement on the perturbed system
will be the vanishing of the first two even derivatives of the radial wave function, that is,
$\partial^2_r\chi(r,t)|_{r=0}=0=\partial^4_r\chi(r,t)|_{r=0}$.

\subsection{Dimensionless version of the system}\label{sec_dimensionless}

In the perturbed SN system \eqref{eq_SNRCspherical}--\eqref{eq_PoissonSpherical} there are many coupling constants and parameters
that obscure the meaning of each term. In order to facilitate the analysis of these equations,
we will perform a nondimensionalization of the system so that the number of parameters and
constants involved is reduced to the minimum required.
First, we will fix the parameters $\beta$ and $\gamma$ to their corresponding values
in the theory of general relativity, that is, $\beta=1=\gamma$.
Next, making use of the physical parameters of the particle $m$ and $\sigma$, in combination
with the Planck constant $\hbar$, it is possible to define the following dimensionless
coordinates:
\begin{equation}
\tau:=\frac{\hbar}{\sigma^2m} t, \quad \quad \rho:=\frac{r}{\sigma}.
\label{nond1}
\end{equation}
Note, in particular, that these
are the only dimensionless coordinates one can construct without making use of the constants
$G$ and $c$, and thus they are also natural coordinates for the nonrelativistic
$(c\rightarrow\infty)$ and the free-particle ($G\rightarrow 0$) cases.
Concerning the dependent variables, $\chi$ and $\Phi$, it is straightforward to define
their dimensionless versions as
\begin{equation}
 S(\rho,\tau):= \sqrt{ \sigma}\ \chi(\rho, \tau), \quad \quad V(\rho, \tau):=\frac{\sigma}{Gm} \Phi(\rho, \tau).
 \label{nond2}
\end{equation}
The interpretation of these functions is clear:
$|S(\rho,\tau)|^2d\rho$ provides the probability of finding the particle between a radius
$\sigma\rho$ and $\sigma(\rho+d\rho)$, while $V$ is the ratio between the potential $\Phi$ and the Newtonian potential generated by a point mass $m$
at a distance $\sigma$.

Finally, the constants $G$ and $c$ will be absorbed by considering the following two
dimensionless coupling
constants:
\begin{align}
\lambda&:=\frac{\hbar^2}{8m^2c^2\sigma^2}=\frac{1}{8}  \left(\frac{m_P}{m}\right)^2 \left(\frac{l_P}{\sigma}\right)^2, \label{eq_lambda}\\
\kappa&:= \frac{Gm^3\sigma}{\hbar^2}=\left(\frac{m}{m_P}\right)^3\left(\frac{\sigma}{l_P}\right), \label{eq_kappa}
\end{align}
where the Planck mass $m_P$ and length $l_P$ have been introduced in order
to have a reference of the magnitude of these quantities.
Performing all the above changes, the relativistic SN system \eqref{eq_SNRCspherical}--\eqref{eq_PoissonSpherical} is rewritten into its dimensionless form as
\begin{align}
&     i \frac{\partial S}{\partial\tau}=-\frac{1}{2}\frac{\partial^2 S}{\partial \rho^2}+\kappa V   S  -\lambda \frac{\partial^4 S}{\partial \rho^4}+4\lambda\kappa^2 V^2 S
     -6 \lambda \kappa \left( \frac{\partial^2  V }{\partial \rho^2}  S +2 \frac{\partial   V}{\partial \rho} \frac{\partial  S}{\partial \rho} +2  V\frac{\partial^2 S}{\partial \rho^2}\right),
    \label{eq_SNRCsphericaldimless}
    \\
&\frac{\partial}{\partial \rho}\left(\rho^2\frac{\partial  V}{\partial \rho}\right)=|S|^2.
\label{eq_PoissonSphericaldimless}
\end{align}
For completeness, we also provide here the initial and boundary conditions we have
commented on in the previous subsection for this dimensionless system:
\begin{equation}\label{initialS}
    S(\rho,0)= \frac{2}{\pi^{1/4}}\ \rho e^{-\frac{\rho^2}{2}},
\end{equation}
\begin{equation}\label{boundaryS}
    S(0, \tau)=0, \quad  S(\infty, \tau)=0, \quad \partial^2_\rho S(\rho,\tau)\Big|_{\rho=0}=0, \quad \partial^4_\rho S(\rho,\tau)\Big|_{\rho=0}=0,
\end{equation}
\begin{equation}\label{boundaryV}
     \partial_\rho V(\rho,\tau)\Big|_{\rho=0}=0, \quad \partial_\rho V(\rho,\tau)\Big|_{\rho=\infty}=-\frac{V(\rho,\tau)}{\rho}\Bigg|_{\rho=\infty}.
\end{equation}

In this way, the only free parameters of the equations
\eqref{eq_SNRCsphericaldimless}--\eqref{eq_PoissonSphericaldimless} are the coupling
constants $\lambda$ and $\kappa$. On the one hand, up to a global numerical factor,
the parameter $\lambda$ is the ratio between the expectation value of the kinetic
energy of a spherical Gaussian wave packet $\langle{\bf p}^2\rangle/(2m)=3\hbar^2/(4m\sigma^2)$ and its rest energy $mc^2$. This coupling constant represents the strength of
the relativistic effects, and $\lambda= 0$ corresponds to the Newtonian ($c\rightarrow\infty$) limit. On the other hand, the parameter $\kappa$ is proportional to the ratio between the expectation values of the Newtonian gravitational energy
$\langle G m^2/r \rangle=2Gm^2/(\sqrt{\pi}\sigma$) and of the kinetic energy $\langle{\bf p}^2\rangle/(2m)$
for a spherical Gaussian wave packet.
Therefore, $\kappa$ encodes the self-gravitational effects and the limit $\kappa=0$
represents the free-particle ($G=0$) case. Note also that these coupling constants contain
information not only of the universal constants $G$, $\hbar$, and $c$, but also
of properties of the particle. In particular, once the mass $m$ of the particle
and its initial width $\sigma$ are chosen, the parameters $\kappa$ and $\lambda$ are fixed. Hence, the two-dimensional parameter space of the model can be equivalently coordinatized
by $(m,\sigma)$ or  $(\lambda,\kappa)$.

It is interesting to note that relativistic and self-gravitational effects
have a different tendency with the mass and the width of the particle: while increasing
$m$ and $\sigma$ increases the value of $\kappa$, it decreases the value of $\lambda$. For instance, to give an idea of the order of magnitude of the different terms
in equation \eqref{eq_SNRCsphericaldimless}, 
for a proton $\kappa\approx 10^{-36}$ and $\lambda\approx 10^{-7}$,
for the He atom $\kappa\approx 10^{-31}$ and $\lambda\approx 10^{-13}$, while,
for a large atom like, for instance, Hg, one gets $\kappa\approx 10^{-25}$
and $\lambda\approx 10^{-18}$,
where we have taken $\sigma$ as the characteristic diameter of each object.
For all these examples, the term $\kappa\lambda$ is approximately of
the same order of magnitude $\kappa\lambda\approx 10^{-43}-10^{-44}$,
and the coefficient $\lambda\kappa^2\propto m^4$ does not depend on $\sigma$ but
simply increases with the mass of the particle.
Hence, as expected, all these terms are extremely small for usual particles
and atoms. However, since our aim is to test numerically their physical effects,
we will need to assume larger values than the commented ones for the coupling constants;
otherwise, the numerical error will completely hide the effects under study.
Therefore, let us first analyze the validity of the model and check
what are the admissible ranges of values for the different parameters.

\subsection{Validity of the model and studied cases}\label{sec_validity}

\begin{figure}[t]
    \centering
    \includegraphics[width=\textwidth]{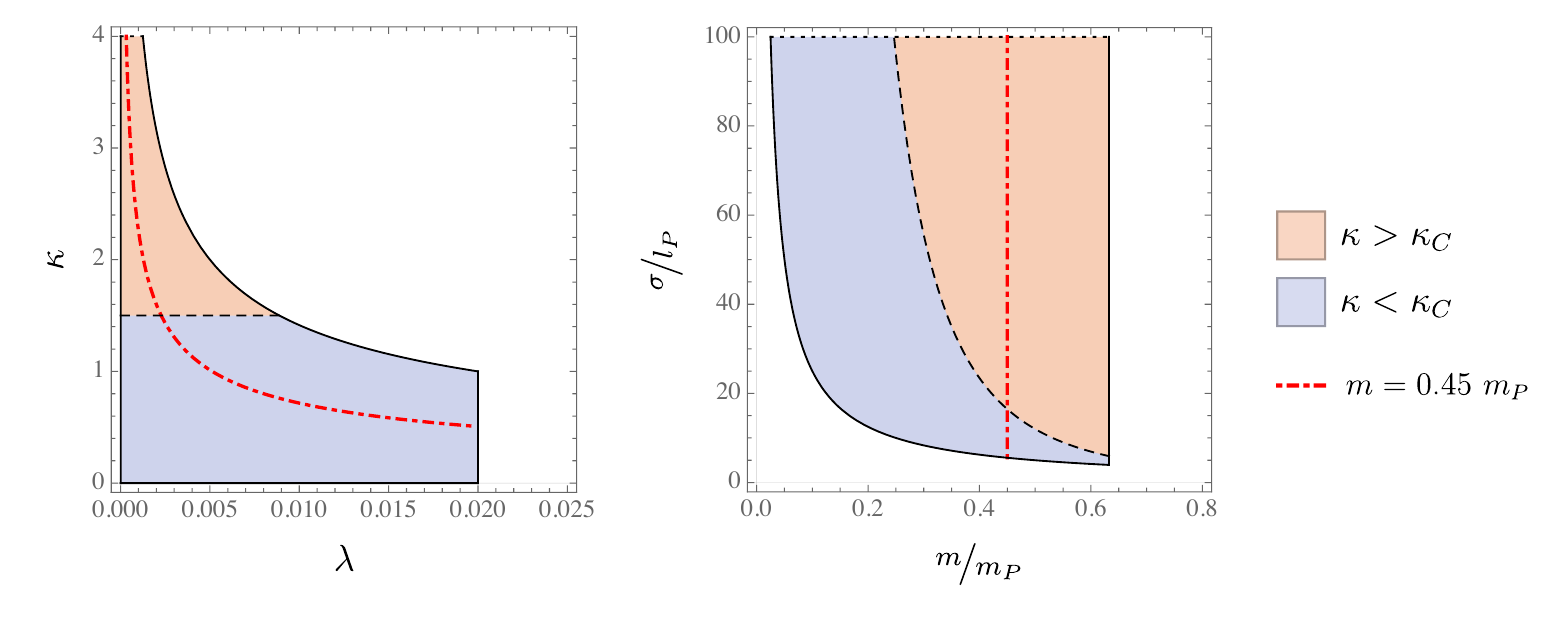}
    \caption{In these plots we show the parameter space of the model, both in the representation given by
    the couple $(\lambda,\kappa)$ and $(m,\sigma)$. The regions where the model is valid (that is, where $\kappa^2 \lambda\lesssim\epsilon$ and $\lambda\lesssim\epsilon$ are obeyed)
    for the particular choice $\epsilon=0.02$ are bounded by black continuous lines. Note that the upper boundary in both graphs is represented as a dotted line, since it is not an actual limit, and the region of possible states continues beyond.
    The allowed regions are colored either in orange or blue, which correspond to states with strong ($\kappa>\kappa_c$) or weak ($\kappa<\kappa_c$) self-gravitational interaction, respectively. The discontinuous black line $\kappa=\kappa_c\approx 1.48$ separates both regimes.
        Finally, the red dot-dashed line shows the particles with fixed mass $m=0.45\, m_P$, which are
        the ones considered for the numerical study
        . }
    \label{fig_possiblestates}
\end{figure}

The model has been constructed by considering a post-Newtonian expansion and
dropping terms of an order higher than $1/c^2$. Thus,
to describe accurately the physical system under consideration, all relativistic correction terms must be small.
Since the functions $S$ and $V$ are dimensionless and normalized (their values are of order $1$),
taking into account the coefficients of the relativistic terms that appear in the SN equation \eqref{eq_SNRCsphericaldimless}, a sensitive choice of bounds for the positive parameters $\lambda$ and $\kappa$ are given by the following three conditions:
$ \kappa^2 \lambda\lesssim\epsilon$, $ \kappa \lambda\lesssim\epsilon$, and $\lambda\lesssim\epsilon$,
with $\epsilon$ being a small positive constant.
It is easy to see that the second condition is redundant, which leaves only two independent bounds: $\kappa^2 \lambda\lesssim\epsilon$ and $\lambda\lesssim\epsilon$.
These conditions can be translated to bounds for the mass $m$ and the initial width $\sigma$ of the particle:
$m\lesssim(8\epsilon)^{1/4}m_P$ and $m\sigma\gtrsim(8\epsilon)^{-1/2}m_Pl_P$.
In particular, this means that, in order to keep the perturbative terms small,
the mass of an initial wave packet is bounded from above. However, note that this bound
is very large: even if this model were valid only up to a very small value of
$\epsilon$, say, for instance, $\epsilon\approx 10^{-20}$, the upper
bound of the mass would be around $10^{-5}m_P$.
In Fig. \ref{fig_possiblestates} the regions of the planes 
($\lambda$, $\kappa$) and ($m$, $\sigma$) satisfying the discussed conditions
are shown for the particular value of $\epsilon=0.02$. Therefore, any system
with values ($\lambda$, $\kappa$), or ($m$, $\sigma$), lying in those regions could be described
by the present model, whereas for a system with values outside these regions,
one would need to consider higher-order post-Newtonian correction terms.

In order to analyze the parameter space in the most efficient way, we will choose a particle of fixed mass
and then change its width $\sigma$ in the range of its allowed values.
Since we want the relativistic effects to be as large as possible
while respecting the bounds given above, the mass will be chosen of
the same order of magnitude as the maximum allowed mass.
For concreteness, if we set $\epsilon=0.02$ for our practical purposes, the conditions
for the mass $m$ and the width $\sigma$ read $m\lesssim 0.63\, m_P$ and
$m \sigma \gtrsim 2.5\, m_P l_P$. Therefore, we will choose a particle with a fixed mass
$m=0.45\, m_P$, and its width $\sigma$ will then need to obey
$\sigma\gtrsim 5.6\, l_P $. Starting from this minimum value, we will
increase $\sigma$ following the red line shown in Fig. \ref{fig_possiblestates}.
For the different values of $(m,\sigma)$, we will numerically solve
the perturbed system \eqref{eq_SNRCsphericaldimless}--\eqref{eq_PoissonSphericaldimless} as well as its background version, which is obtained just by setting $\lambda=0$ in \eqref{eq_SNRCsphericaldimless}, and compare the differences
in the evolution of the wave function. These differences will thus be the relativistic
1PN effects. The technical details about the numerical methods are presented in Appendix A.

\subsection{Results and discussion}\label{sec_results}

Since the relativistic effects considered in our numerical study are
small, the observed evolution of the system is quite similar for both relativistic
and nonrelativistic systems. Let us thus begin this section by providing a
qualitative description of the dynamics, which applies to both cases.

Note that in Fig. \ref{fig_possiblestates} the validity regions of the model commented above
have been colored, either in orange or blue, depending on whether the corresponding $\kappa$
is larger or smaller than certain value $\kappa_c$. This critical value separates
two qualitatively distinct evolutions of the wave packet.
On the one hand, small values of $\kappa$ define
the weak self-gravitational regime, where the gravitation is not large enough to counteract the natural dispersion of the wave function. In this case, the wave function evolves in a similar
fashion as in
the free-particle case, but slower due to the gravitational attraction.
On the other hand, one can define the strong self-gravitational regime, where self-gravitation fully compensates the quantum spreading of the wave function. In this regime, the wave packet oscillates around an equilibrium point, and it eventually decays
to a stationary state. During this process, known as gravitational cooling, part of the probability is slowly
radiated away to infinity, while the amplitude of the oscillations decreases until the wave function becomes completely stationary \cite{Harrison2002}.
The transition between these two regimes is quite sharp and approximately happens
at the value $\kappa_c \approx 1.48$.
This precise value for $\kappa_c$
has been obtained in numerical analyses of the nonrelativistic SN equation \cite{VanMeter2011}.
However, one can provide a heuristic derivation, based on the Newtonian
escape velocity, which can be helpful for understanding the underlying physics.

Let us define the velocity of the particle as $v:=\sqrt{\langle {\bf p}^2/m^2 \rangle}$.
For the initial state \eqref{eq_initialstate}, it is then straightforward to obtain
the corresponding initial velocity
$v=\sqrt{3/2}\hbar/(m\sigma)$. Further, the escape velocity for a particle
located at a distance $\alpha\sigma$ from a central mass $M$ is given by
$(2 G M/(\alpha\sigma))^{1/2}$. In the SN model,
this central mass and the position of the particle are related with $M$ being the mass
located in the interior of the sphere of radius $\alpha\sigma$, that is,
$M=m\int_0^{\alpha\sigma}|\chi|^2 dr$, which, for the initial state, takes
the explicit form $M=m({\rm erf}(\alpha)-\frac{2}{\sqrt{\pi}}\alpha e^{-\alpha^2})$.
The escape velocity is thus a function of $\alpha$, and
it is possible to see that it reaches its maximum value
at $\alpha\approx1.51$.
Requesting then that this maximum escape velocity be equal to the initial
velocity, and taking into account the definition \eqref{eq_kappa},
one obtains the critical value for $\kappa=\kappa_c\approx 1.43$, which
approximates very well the numerical value mentioned above. 
Therefore, for $\kappa<\kappa_c$ the initial velocity exceeds the local
escape velocity at all radii $\alpha\sigma$. Following this interpretation, in
this weak self-gravitational regime, the particle can escape to infinity no matter
what is its actual initial position. Thus, the wave function will disperse tending,
to a flat curve, which will imply a decrease of its kinetic energy $\langle{\bf p}^2/(2m)\rangle$
and an increase on the uncertainty of the position.

Concerning the strong self-gravitational regime, as already commented above,
the wave function tends to a stationary state. However, there is a particular value $\kappa_{eq}$
for which the peak (the maximum of the norm of the wave function) of the initial state is already at its equilibrium position. Let us obtain an estimation
for this value. In the stationary state the virial theorem should be obeyed, which, for the
nonrelativistic SN system, reads $2K-Q=0$ with $K:=\langle {\bf p}^2/(2m)\rangle$
and $Q:=\langle m r \frac{d\Phi}{dr} \rangle$. Computing these expectation values
for the initial state \eqref{eq_initialstate}, it is easy to find that $K=3\hbar^2/(4 m\sigma^2)$
and $Q=G m^2/(\sqrt{2\pi}\sigma)$. Therefore, considering again the definition \eqref{eq_kappa},
the virial theorem is obeyed by
the initial Gaussian wave packet if $\kappa=\kappa_{eq}:=3\sqrt{\pi/2}\approx 3.76$.
In this case the peak simply
oscillates around its initial position $\rho=1$.
The equilibrium value $\kappa_{eq}$ separates two qualitative behaviors.
On the one hand, if $\kappa_c<\kappa<\kappa_{\rm eq}$, the kinetic energy dominates
$(2K-Q>0)$, and the system begins to expand by moving the peak of the wave function
to larger radii. After expanding, it will start oscillating around its equilibrium radius
$\rho>1$. On the other hand, if $\kappa_{\rm eq}<\kappa$, the potential term
dominates $(2K-Q<0)$ and the initial evolution of the system is a collapse
to smaller radii. In this latter case, the peak of the wave function will end up oscillating
around certain $\rho<1$.

As already commented above, the described qualitative picture is unchanged when considering
small relativistic corrections. Now, in order to analyze and discuss the specific effects produced
by the relativistic terms, we will first consider the weak and afterwards the strong
self-gravitational regimes. At this point it is important to recall that all the presented
simulations correspond to a particle with fixed mass $m=0.45\, m_P$ and varying width $\sigma>3.5\, l_P$.
Note that, for this set of states,
increasing $\sigma$ corresponds to increasing $\kappa$ \eqref{eq_kappa}
and to decreasing $\lambda$ \eqref{eq_lambda}. In fact, from the above value for $\kappa_c$, one can define the critical value
$\sigma_c\approx 16.2\, l_P$, so that narrow states $(3.5\, l_P<\sigma<\sigma_c)$
are in the weak self-gravitational regime, while wide states $(\sigma_c<\sigma)$
follow the strong self-gravitational behavior.

\subsubsection{Weak self-gravitational regime $(\kappa<\kappa_c, \,\sigma<\sigma_c)$}

\begin{figure}[t]
    \centering
    \includegraphics[width=\textwidth, keepaspectratio=false]{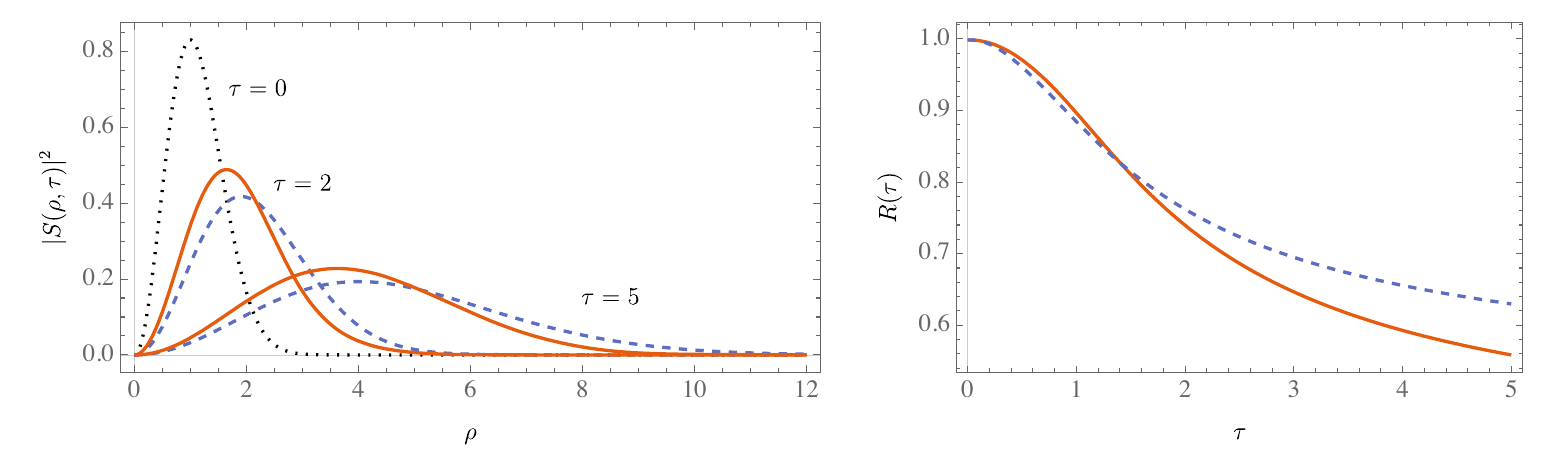}
    \caption{The plot on the left shows the profile of the probability distribution $|S|^2$ at
    different times for the evolution
    given by the relativistic (orange solid line) and nonrelativistic (blue dashed line) SN systems.
    The initial state at $\tau=0$ is the same for both cases, a
    spherical Gaussian wave packet with mass $m=0.45\ m_P$ and width $\sigma=10\ l_P$, and its corresponding
    profile is represented by the black dotted line. The plot on the right corresponds to
    the evolution of the ratio between the expectation value of the kinetic energy and the kinetic energy of the initial state $R(\tau)$ for
    the relativistic (orange solid line) and nonrelativistic (blue dashed line) cases.}
    \label{fig_lowregime}
\end{figure}

Concerning the nonrelativistic evolution of the wave function, in this weak self-gravitational
regime, 
and in accordance with other studies in the literature (see, e.g., Refs.\cite{VanMeter2011,Giulini2011, Silvestrini2015}), we observe that 
the initial state evolves qualitatively as a free particle,
except for the dispersion velocity, which is slightly lower. Considering the evolution of the same initial wave packet
under the relativistic system \eqref{eq_SNRCsphericaldimless}--\eqref{eq_PoissonSphericaldimless},
the wave function is observed to be spreading even slower,
while still keeping the dispersive tendency.
In order to show the differences between the relativistic and nonrelativistic cases,
in Fig. \ref{fig_lowregime} the profile of the probability distribution $|S|^2$
is plotted at different times. As time passes by, in both cases the bell shape
of the probability distribution is kept, while the peak (the maximum of $|S|^2$), initially located at $\rho=1$,
moves to larger radii and its width (standard deviation) increases. However,
it is clearly seen that in the relativistic case this dispersive process is slower.

In addition, in Fig. \ref{fig_lowregime} the evolution of the ratio between the expectation
value of the kinetic energy and the kinetic energy of the initial state
$R(\tau):=\frac{2\sigma^2}{3\hbar^2}\langle {\bf p}^2 \rangle =-\frac{2}{3}\int_0^\infty  d\rho \overline{S} \frac{\partial^2 S}{\partial\rho^2}$ is shown.
In both relativistic and nonrelativistic cases, the expectation value of the kinetic energy is maximum
for the initial state and it decreases, as evolution goes on, tending to zero in the
$\tau\rightarrow\infty$ limit. But this decrease is observed to be faster
in the relativistic case. Therefore, one can state that in this regime
the relativistic corrections slow down the dynamics of the wave function.

\subsubsection{Strong self-gravitational regime $(\kappa>\kappa_c,\,\sigma>\sigma_c)$}

In order to describe the dynamics in this regime, we will consider the evolution
of the position of the peak (the maximum of the module of the wave function) $\rho_{\text{peak}}(\tau)$. This function follows a damped oscillatory behavior, slowly tending to its equilibrium value in the limit $\tau\rightarrow\infty$,
and it encodes the dynamical information about the approach of the wave function to its final stationary state.
In particular, it is very convenient to define its mean value as
$\bar{\rho}_{\text{peak}}:=1/(\tau_f-\tau_i)\int_{\tau_i}^{\tau_f}\rho_\text{peak}(\tau)d\tau$,
with $\tau_i$ and $\tau_f$ being the initial and final values of $\tau$ for
the corresponding numerical evolution, respectively. This mean value is not the exact final equilibrium
point, since the oscillations are not completely regular. However, by performing long numerical
evolutions (in our case, we have considered up to $\tau_f-\tau_i=500$), it provides a very good approximation
of the equilibrium position.
In addition, the irregularity of the oscillations also makes it difficult to calculate
its frequency (it is not constant, it rather changes over time),
but it is possible to discuss it at a qualitative level.

Regarding the nonrelativistic evolution in the strong self-gravitational regime, the results we have
obtained are in complete agreement with the ones presented in the literature
(see, e.g., Refs. \cite{Guzman2004, Harrison2002, VanMeter2011, Giulini2011, Guzman2006, Silvestrini2015, Manfredi2013}).
As already described above,
the position of the peak $\rho_{\rm peak}$ oscillates around a fixed radius, while the amplitude of these oscillations
decreases, following a pattern of gravitational cooling. The amplitude of these oscillations depends on
the specific width of the state $\sigma$, and thus on the value of $\kappa$. In particular, it is
observed to be minimum  around $\sigma\approx 38\ l_P$, which corresponds to $\kappa\approx3.5$,
and it increases for both wider and narrower states. Note that this value
is very close to the $\kappa_{\rm eq}$ derived above making use of the virial theorem, and thus
represents the case where the initial position of the peak is already at its equilibrium point,
which explains the small amplitude of the oscillations.

\begin{figure}[H]
    \centering
    \begin{subfigure}[l]{\textwidth}
        \centering
        \includegraphics[width=0.7\textwidth]{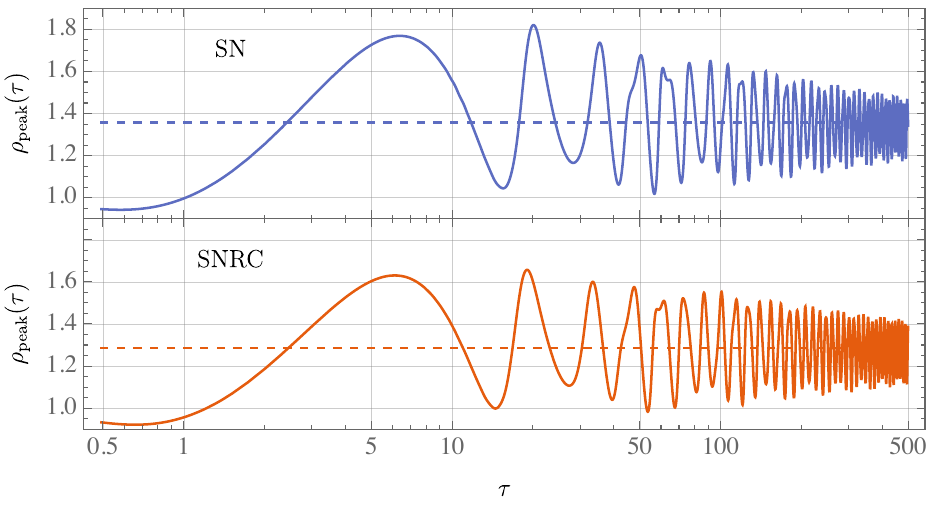}
        \caption{$m=0.45\ m_P,\ \sigma=30\ l_P \quad \Longrightarrow \quad \lambda=0.0007,\  \kappa=2.73 <\kappa_{\rm eq}$.}
    \end{subfigure}
    \begin{subfigure}[l]{\textwidth}
        \centering
        \includegraphics[width=0.7\textwidth]{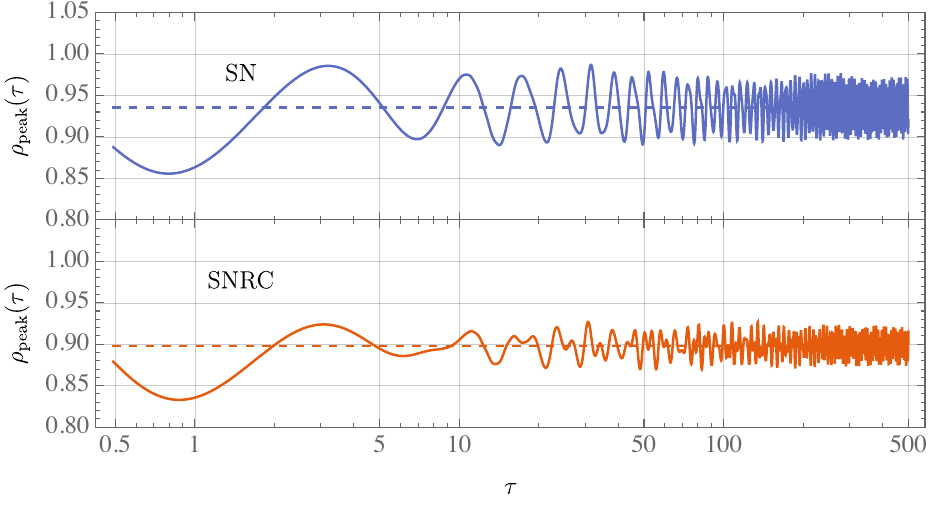}
        \caption{$m=0.45\ m_P,\ \sigma=40\ l_P \quad \Longrightarrow \quad\lambda=0.0004,\   \kappa=3.65\approx\kappa_{\rm eq} $.}
    \end{subfigure}
    \begin{subfigure}[l]{\textwidth}
        \centering
        \includegraphics[width=0.7\textwidth]{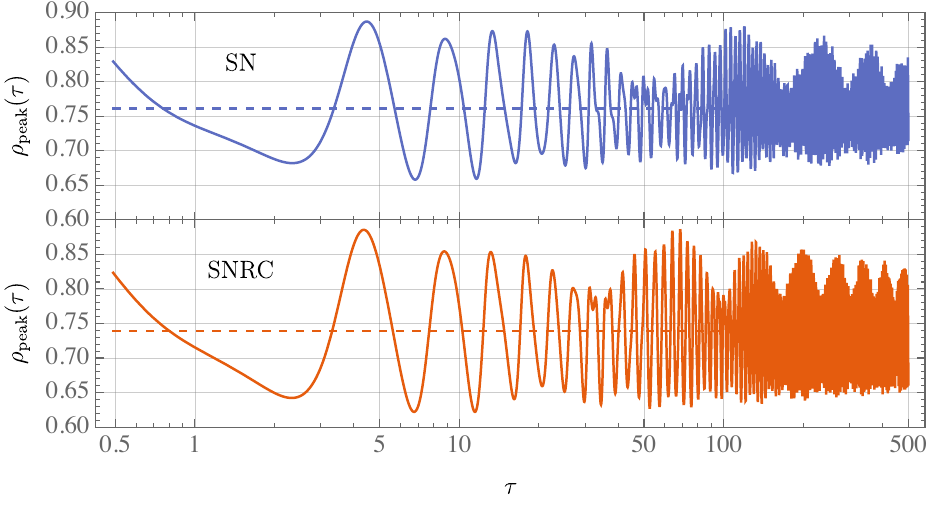}
        \caption{$m=0.45\ m_P,\ \sigma=50\ l_P \quad \Longrightarrow \quad \lambda=0.0002,\ \kappa=4.56>\kappa_{\rm eq}$.}
    \end{subfigure}
\caption{Evolution of the position of the peak of the wave function $\rho_{\text{peak}}(\tau)$ for three different initial states. In each case, the evolution under the nonrelativistic SN equation is shown on the top with a blue curve. On the bottom, with an orange curve, the evolution given by the SN equation with relativistic corrections (SNRC) is depicted. The
horizontal dashed lines represent the corresponding mean value of the position $\bar{\rho}_\text{peak}$.}
\label{StrongRegime}
\end{figure}

As can be seen
in Fig. \ref{StrongRegime}, where the nonrelativistic and relativistic evolution of the
peak of the wave function is shown for some characteristic examples,
the evolution of the same initial states under the SN equation with relativistic corrections \eqref{eq_SNRCsphericaldimless}--\eqref{eq_PoissonSphericaldimless}
leads to the same gravitational cooling pattern.
However, there are three main effects produced by the relativistic terms.
First, regardless the value of $\kappa$, in all the studied cases we observe a reduction of around $5\%$ of the mean value of the position of the peak $\bar{\rho}_{\rm peak}$
as compared to the nonrelativistic evolution.
For instance, in the particular cases depicted in Fig. \ref{StrongRegime}, the mean value of the peak is reduced from $\bar{\rho}_\text{peak}=1.36$ to $\bar{\rho}_\text{peak}=1.26$ in the case with {$\sigma=30\ l_P$}, from
$\bar{\rho}_\text{peak}=0.94$ to $\bar{\rho}_\text{peak}=0.90$ for {$\sigma=40\ l_P$},
and from $\bar{\rho}_\text{peak}=0.76$ to $\bar{\rho}_\text{peak}=0.74$ for {$\sigma=50\ l_P$}. This means that the oscillations are taking place around a smaller radius. In fact, the peak of the final stationary
state will be approximately located at this position and, thus, we conclude that the relativistic
corrections produce a more compact stationary state.
Second, as can be seen in the plots shown in Fig. \ref{StrongRegime},
the amplitude of the oscillations is modified:
for $\kappa\lesssim \kappa_{\rm eq}$ the amplitude decreases when considering relativistic effects, while for
$\kappa\gtrsim \kappa_{\rm eq}$ it increases.
The third feature concerns the frequency of the oscillations. Excluding
the states with a value of $\kappa$ around $ \kappa_{\rm eq}$, the frequency of the oscillations increases
when relativistic corrections are taken into account.
Furthermore, and particularly for very large values of $\kappa$,
the envelope of the oscillations also shows a larger frequency for the relativistic case.
Therefore, in this sense, the relativistic system evolves more rapidly than
the nonrelativistic one.

In order to interpret these effects,
let us make an analogy with the much simpler system given by a damped harmonic oscillator.
On the one hand,
in such a system the amplitude of the oscillations is proportional to the initial deformation
and it is exponentially modulated in terms of the corresponding friction parameter.
Therefore, the greater the initial deformation, the larger the amplitude of the oscillations.
In our system this initial deformation can be defined as the difference between the equilibrium
point and the initial position of the peak $\rho_{\text{peak}}(0)=1$. Since, as commented above,
the equilibrium point for the relativistic case is always smaller than in the nonrelativistic case,
the corresponding initial deformations are different, which explains the observed behavior.
More precisely, for $\kappa\lesssim\kappa_{\rm eq}$, the initial deformation for the relativistic
system is smaller than for the nonrelativistic system, leading to smaller oscillations.
For $\kappa\gtrsim \kappa_{\rm eq}$ just the opposite occurs:
a larger initial relativistic deformation produces oscillations with a larger amplitude.
On the other hand, if the harmonic oscillator is driven by an external gravitational
field, its frequency is proportional to the square root of the local gravitational acceleration,
and thus, the increase of such acceleration would produce an enhancement of the frequency.
In conclusion,
all these effects are in complete accordance with the interpretation of the relativistic terms
effectively producing a stronger self-gravitational interaction.

\section{Conclusions}

In this article we have presented a modification of the Schr\"odinger-Newton
equation considering certain relativistic corrections to test whether this model
is still a valid approach to explain the localization of the wave function
when relativistic effects are not negligible.
For such a purpose, we have started from the Hamiltonian of a particle propagating
on a curved background.
Then, making use of the parametrized post-Newtonian formalism and assuming spherical symmetry for the background, we have performed
an expansion of the Hamiltonian in inverse powers of the speed of light up to
the 1PN order, that is, up to the order $c^{-2}$.
In order to obtain a simple framework, as a first approximation
to the complete picture, we have only considered the terms given by the
Newtonian potential, and thus other post-Newtonian potentials have been
neglected.
Next, by promoting the basic variables and their conjugate momenta to operators,
we have carried out the canonical quantization of the system.
At this point, the gravitational potential appears as a free function in
the corresponding Schr\"odinger equation. Therefore, imposing that there is no other
source to the potential than the self-gravitation of the particle itself,
and as it is done
in the usual nonrelativistic model, a mass density has been defined
in terms of the square of the module of the wave function. This mass density sources
the Poisson equation, which, in turn, defines the gravitational potential.
This procedure has completed the construction of the Schr\"odinger-Newton
model with relativistic corrections up to the 1PN order. The dynamics
of the wave function and the gravitational potential is thus described by the
nonlinear coupled system of differential equations \eqref{eq_snrc}--\eqref{eq_snrcpoisson}.

Due to the complexity of those equations, we have studied the dynamical evolution of initial
Gaussian wave packets using numerical techniques. This
initial state is completely characterized by two parameters: the mass of the particle $m$
and the Gaussian width $\sigma$. It is interesting to note that, making use of these parameters,
in combination with universal constants, it is possible to obtain the dimensionless version
of the system \eqref{eq_SNRCsphericaldimless}--\eqref{eq_PoissonSphericaldimless}. As can be seen, written in this form, there are only two
dimensionless coupling constants, $\kappa$ and $\lambda$, which respectively weight the strength
of the self-gravitational and the relativistic effects. In order to make the relativistic effects
larger than the numerical error while still keeping them small enough that they can be
interpreted as perturbations, we have chosen a relatively large mass of the particle. For concreteness,
this mass has been kept fixed for the different numerical simulations, while the
initial width of the state has been varied. This has allowed us to explore distinct
interesting regions of the parameter space.

In order to get a clear picture about how relativistic corrections influence the dynamics, we have
compared the evolution of the chosen states under both the relativistic and the nonrelativistic Schr\"odinger-Newton equations.
As is well known from previous analysis of the nonrelativistic system,
two regimes can be distinguished depending on the value of the parameter $\kappa$ or,
equivalently for the set of (fixed-mass) states we have considered, on the initial width $\sigma$.
In the weak self-gravitational regime,
the wave packet disperses as for the free-particle case, but more slowly.
In contrast, in the strong self-gravitational regime,
the wave packet oscillates around a certain fixed radius, slowly decaying into a stationary state.
On the one hand, we have observed that, in the weak self-gravitational regime, the dispersion of the wave packet is slower under the presence of relativistic correction terms. On the other hand, in the strong self-gravitational regime, the most relevant effect of the relativistic terms is a reduction of the equilibrium radius
around which the peak of the wave function oscillates. In particular,
this behavior indicates that the wave function will eventually settle into a more compact stationary state.
In addition, the frequency of the oscillations is increased,
producing, in this sense, a faster evolution of the relativistic system, while the change in the
amplitude depends on the specific value of the initial width.
All in all, one can state that relativistic terms effectively increase
the self-gravitation of the particle and thus reinforce the validity of the Schr\"odinger-Newton
model as an explanation for the localization of quantum states by gravitational collapse.
A similar result was found in Ref. \cite{Manfredi15}, where the relativistic effects
produced in this system by the gravitomagnetic vector potential were studied.

\section*{Acknowledgments}
DB thanks Leonardo Chataignier and Raül Vera for interesting discussions,
Giovanni Manfredi for correspondence, and
the Max-Planck-Institut f\"ur Gravitationsphysik (Albert-Einstein-Institut)
for hospitality while part of this work has been done.
This work is supported by the Basque Government Grant \mbox{No.~IT1628-22},
by the Grant PID2021-123226NB-I00 (funded by MCIN/AEI/10.13039/501100011033 and by “ERDF A way of making Europe”),
and by the Alexander von Humboldt Foundation.

\appendix

\section{Numerical method}

In order to obtain a numerical solution to the nonlinear coupled system of equations,
\eqref{eq_SNRCsphericaldimless} -- \eqref{eq_PoissonSphericaldimless}
we have used the algebraic computational software \textit{Mathematica}. In this code
the module \mtica{NDSolve}, included in the usual distribution, is the standard numerical solver.
However, it is unable to solve such complicated systems without previous simplifications.
Therefore, we have implemented the numerical method of lines. This method consists
in reducing a partial differential equation to a coupled system of $N$ ordinary differential equations.
The method has been applied as follows. First, we have split the wave function into
its real and imaginary parts as $S(\rho,\tau)=F(\rho,\tau)+i\, I(\rho,\tau)$. In this way, Eq. \eqref{eq_SNRCsphericaldimless} has been divided into its real and imaginary parts as well.
The gravitational potential $V(\rho,\tau)$ is real, and thus no such decomposition is necessary.
Next, an evenly distributed grid for the spatial coordinate $\rho_n=n\,\Delta\rho$ with $n=0,\dots,N$
and a fixed $\Delta\rho$ have been defined. At each point of the grid, one can then define
the three functions $F_n(\tau):=F(\rho_n,\tau)$, $I_n(\tau):=I(\rho_n,\tau)$, and $V_n(\tau):=V(\rho_n,\tau)$,
which only depend on $\tau$. Finally, the different partial differential equations
have been rewritten in a finite difference approach at
each point of the grid making use of the \mtica{NDSolve`FiniteDifferenceDerivative} extension
of \textit{Mathematica}, which replaces each derivative by the best finite
difference approach up to the desired order of accuracy. 
The boundary conditions and the initial
conditions have also been discretized using \mtica{NDSolve`FiniteDifferenceDerivative}.
This completes the procedure and replaces the coupled system of partial differential equations
\eqref{eq_SNRCsphericaldimless} -- \eqref{eq_PoissonSphericaldimless}
by an approximate coupled system of $3N$ ordinary differential equations in $\tau$.

\begin{figure}[t]
    \centering
    \includegraphics[width=0.95\textwidth]{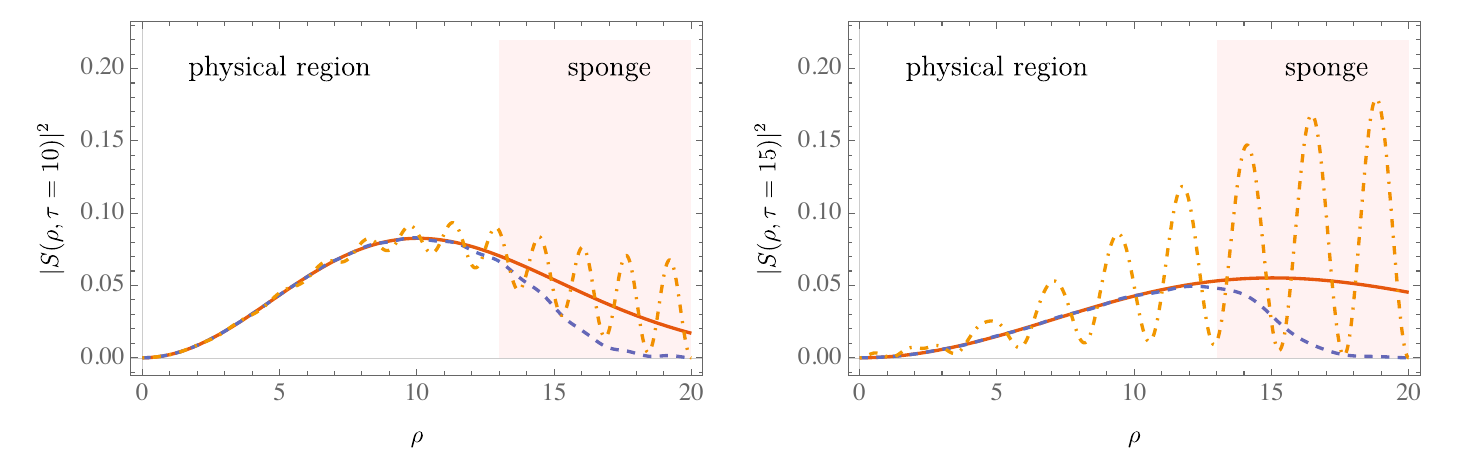}
    \caption{The probability density for a free particle is depicted for different values of time. The analytical solution (orange solid line), the numerical solution with the CAP (blue dashed line), and the numerical solution without the CAP (yellow dot-dashed line) are plotted. }
    \label{fig_sponge}
\end{figure}

Furthermore, one needs to take into account that the numerical solution can only be obtained in a finite domain of the variables $\tau$ and $\rho$. Hence, to avoid reflections on the numerical boundary
$\rho=\rho_N$, we have used a complex absorbing potential (CAP) as in other similar studies
(see, for instance, Refs. \cite{Harrison2002, Guzman2004, Guzman2006}). The CAP method is based on adding an imaginary dispersive
term on the considered Schrödinger-like equation. This term is expected to disperse the wave function in a limited region near the numerical boundary (the so-called sponge), leaving the dynamics on the remaining domain (physical region) unchanged. For this reason, it has to be dominant in the sponge, while being
negligible in the physical region. A widely used CAP is given by
\begin{equation}
    i \frac{V_0}{2}\tanh{\left( \frac{\rho-\rho_c}{\xi} \right)},
\end{equation}
where $V_0$ provides a reference scale for this term, $\rho_c$ is the critical dimensionless radius determining the limit of the physical region, and $\xi$ is related to the smoothness of the transition. These three parameters are to be optimized for each model.

\begin{figure}[t]
    \centering
    \includegraphics[width=0.95\textwidth]{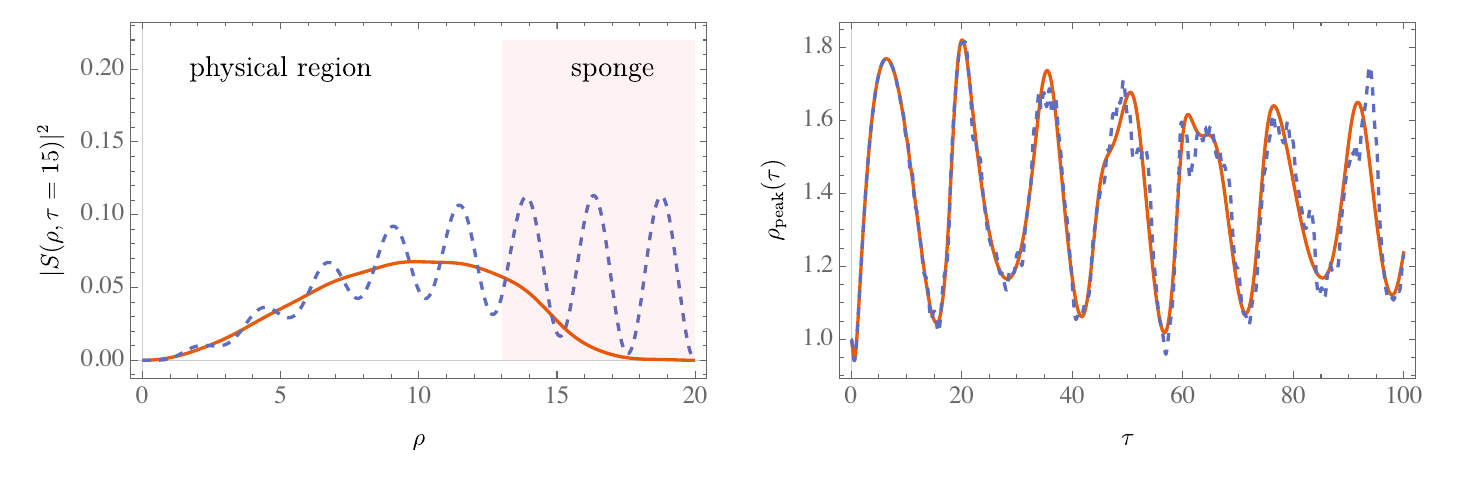}
    \caption{On the left-hand side,
    the probability density of a spherical Gaussian wave packet of width $\sigma=10\ l_P$ under the SN equation with relativistic corrections is shown at a time $\tau=15$, when the wave packet hits the numerical boundary.
    On the right-hand side, the evolution of the peak of a spherical Gaussian wave packet of width $\sigma=30\ l_P$ under the SN equation with relativistic corrections is depicted. In both cases, the solid orange line corresponds to the solution with the CAP and the dashed blue line to the solution without the CAP. }
    \label{fig_spongeSN}
\end{figure}

In this project, different values for the parameters have been tested, comparing the numerical and the exact analytical solutions to the Schr\"odinger equation for a free particle. In this case, the optimal parameters have been found to be $V_0=1$, $\rho_c=16$, and $\xi=3$. With these values, the numerical solution reproduces the analytical solution exactly, while without the CAP, the reflection on the numerical boundary induces spurious oscillations on the result. This comparison can be seen in Fig. \ref{fig_sponge}. In addition, the same values for the parameters
have been tested for the Schr\"odinger-Newton equation with relativistic corrections. Although in this case the numerical solution could not be compared to an analytical solution, when using the CAP, a reduction of some small oscillations originating in the boundary is observed. The comparison of the dynamics with and without
the CAP in the weak and strong self-gravitational regimes is illustrated in Fig. \ref{fig_spongeSN}.

Once the spurious numerical boundary effects are under control, the command \mtica{NDSolve}
has been used to solve the system of ordinary differential equations for each $F_n(\tau)$, $I_n(\tau)$, and $V_n(\tau)$. Finally, the solutions for each point of the spatial grid have been
combined to produce the final numerical results $S(\rho,\tau)$ and $V(\rho,\tau)$ using interpolation.

\section{Alternative modifications of the Poisson equation}

\begin{figure}[t]
    \centering
    \includegraphics[width=0.6\textwidth]{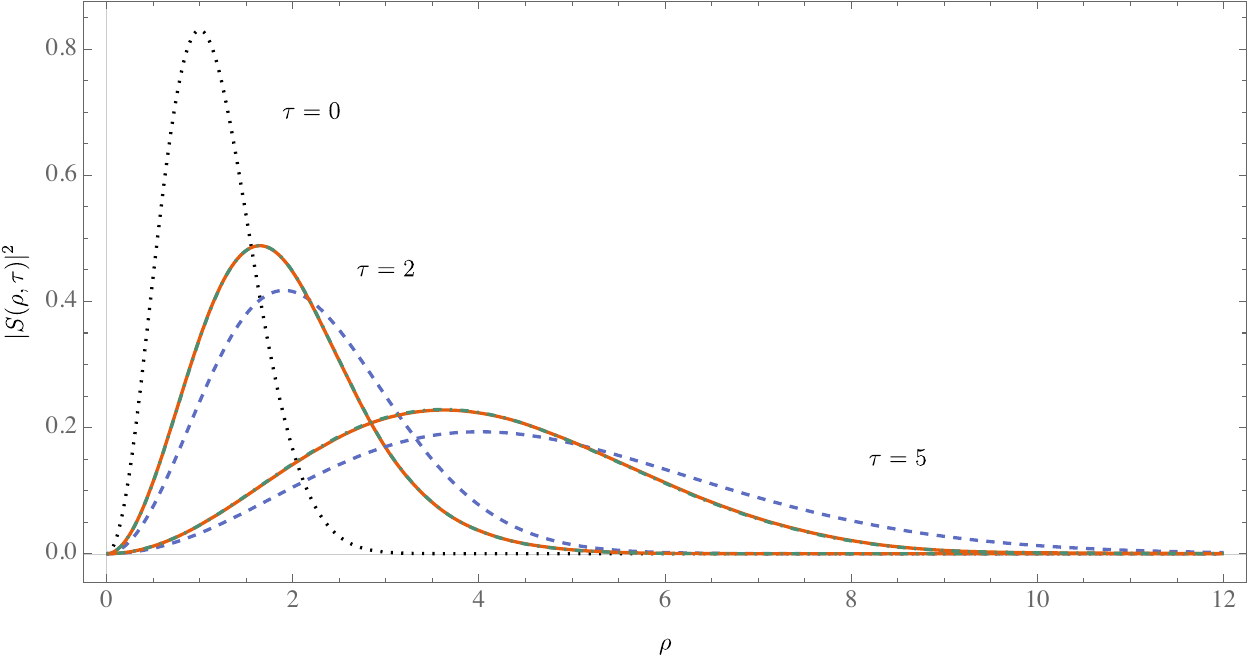}
    \caption{The plot shows the profile of the probability distribution $|S|^2$ at
    different times for the evolution given by the nonrelativistic SN system (blue dashed line)
and the relativistic SN system with the usual Poisson equation (orange solid line).
The green dot-dashed line represents the evolution of the relativistic SN system with the modified version of the Poisson equation \eqref{modPoisson}. Note that the difference between the two relativistic evolutions
is extremely small. The initial state at $\tau=0$ is the same for all cases, a spherical Gaussian wave packet with mass $m=0.45\ m_P$ and width $\sigma=10\ l_P$, and its corresponding profile is represented by the black dotted line.}
    \label{fig_mod1a}
\end{figure}

As already mentioned in the text, in order to take into account both mass and energy densities as gravitational sources,
some studies \cite{Franklin15, Franklin16} proposed a modified Poisson equation that, at 1PN order,
reads $$\Delta\Phi=4\pi G\rho\left(1+\frac{\Phi}{c^2}\right)+\frac{(\nabla\Phi)^2}{2 c^2}.$$
  In this appendix we numerically study whether these modifications affect the results of our model.
  First, using the definitions \eqref{nond1}--\eqref{eq_kappa}, we rewrite this modified Poisson equation into its dimensionless form
\begin{equation}
    \frac{\partial}{\partial \rho}\left(\rho^2\frac{\partial V}{\partial \rho}\right)=|S|^2+8\kappa \lambda|S|^2 V + 4\kappa \lambda \rho^2 \left(\frac{\partial V}{\partial \rho}\right)^2.
    \label{modPoisson}
\end{equation}
Hence, the new system under study is given by the two equations \eqref{eq_SNRCsphericaldimless} and \eqref{modPoisson}, along with the initial condition \eqref{initialS} and the boundary conditions
\eqref{boundaryS}--\eqref{boundaryV}.

\begin{figure}[H]
    \begin{subfigure}[l]{0.49\textwidth}
        \centering
        \includegraphics[width=0.9\textwidth]{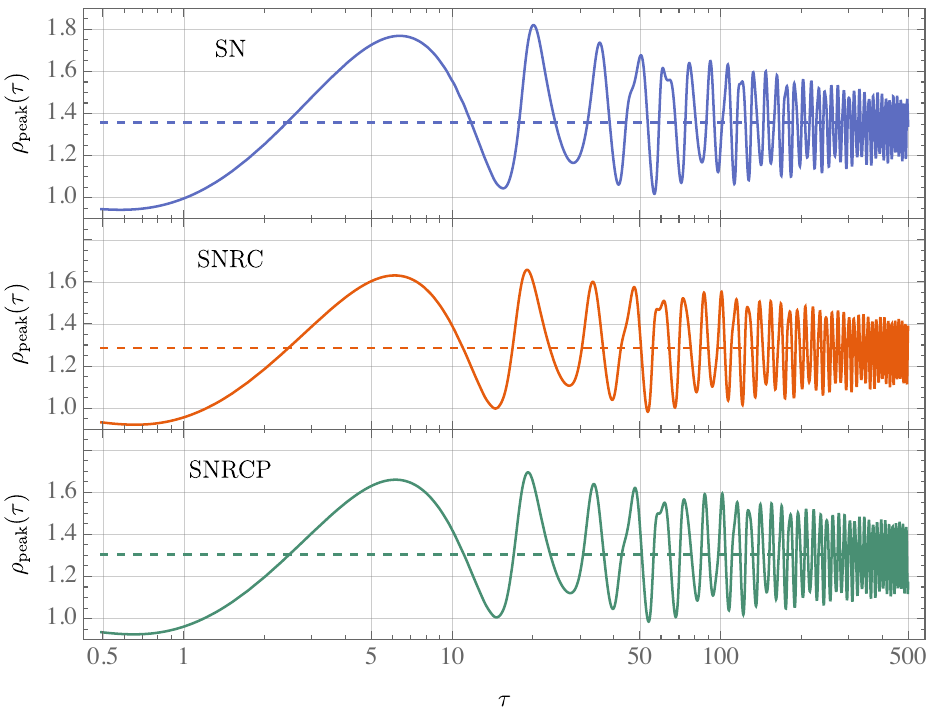}
        \centering
        \caption{$m=0.45\ m_P,\ \sigma=30\ l_P$.}
    \end{subfigure}
    \begin{subfigure}[r]{0.49\textwidth}
        \centering
        \includegraphics[width=0.9\textwidth]{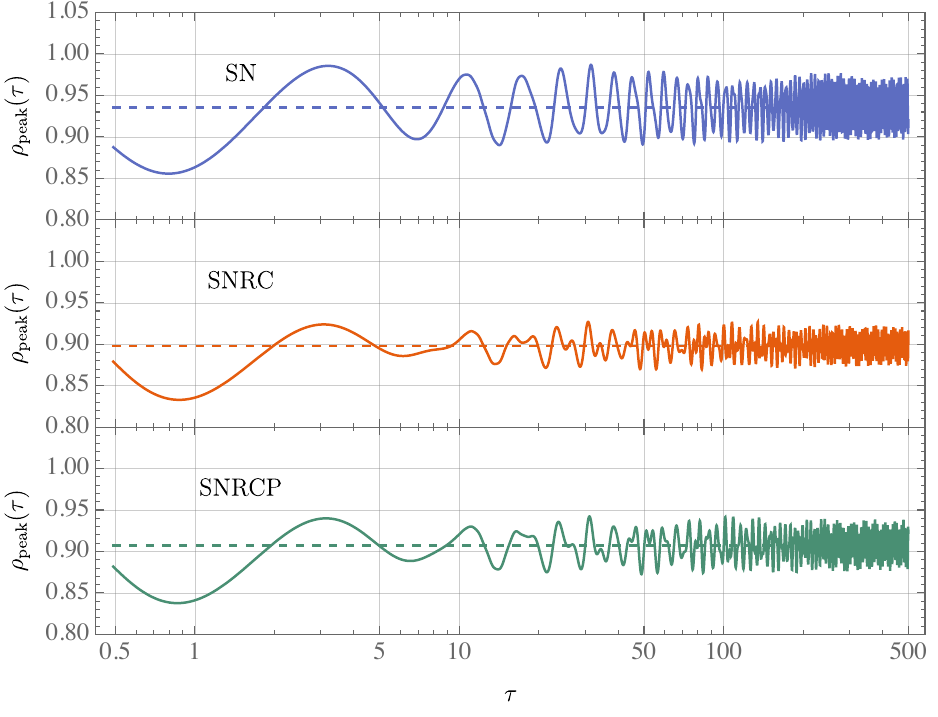}
         \caption{$m=0.45\ m_P,\ \sigma=40\ l_P$.}
    \end{subfigure}
   \par\bigskip
    \centering
    \begin{subfigure}[c]{0.46\textwidth}
        \centering
        \includegraphics[width=\textwidth]{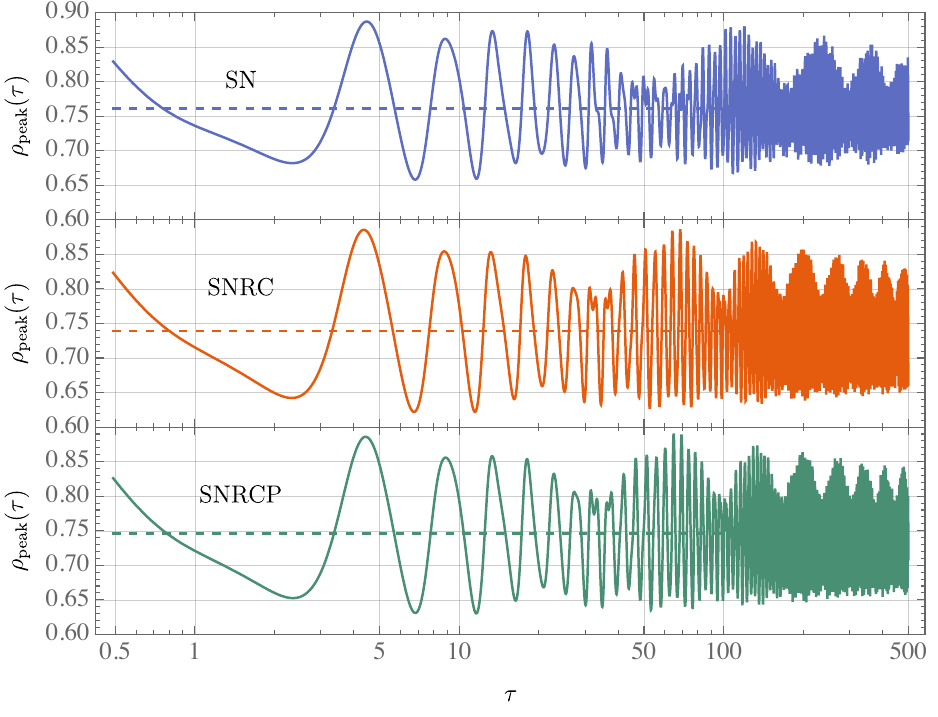}
        \caption{$m=0.45\ m_P,\ \sigma=50\ l_P$.}
    \end{subfigure}
\caption{Evolution of the position of the peak of the wave function $\rho_{\text{peak}}(\tau)$ for three different initial states. In each case, the evolution under the nonrelativistic SN equation
is shown on the top with a blue curve. In the middle, with an orange curve, the evolution given by the SN equation with relativistic corrections (SNRC) is depicted considering the standard Poisson equation.
Finally, on the bottom, a green curve describes the dynamics of the SN equation with relativistic corrections where the gravitational potential is extracted from the modified Poisson equation \eqref{modPoisson} (SNRCP).
The horizontal dashed lines represent the mean value of the position $\bar{\rho}_\text{peak}$
for each evolution.\label{mod2}}
\end{figure}

The dynamics of a wave packet under the modified Poisson equation is qualitatively similar to
the one given by the usual Poisson equation. In fact, in the weak self-gravitational regime
the difference is inappreciable. The wave function spreads more slowly than for the nonrelativistic SN system,
but equally as fast as in the relativistic case (see Fig. \ref{fig_mod1a}). Conversely, in the strong self-gravitational regime, there are some small differences between the results, but they are qualitatively identical, as can be seen in Fig. \ref{mod2}. The main feature that stands out is that the mean value
of the position of the peak of the wave function increases slightly. However, despite this growth, this
mean value is still smaller than for the nonrelativistic case, and thus the interpretation that the relativistic
effects effectively produce a stronger self-gravitation stands.

\end{document}